\documentclass[letterpaper, 10 pt, conference]{IEEEtran}  

\usepackage{balance}  

\usepackage{epsfig,endnotes}
\usepackage{multirow}
\usepackage{url}
\usepackage{diagbox}
\usepackage{enumitem}
\usepackage{lipsum}
\usepackage{graphicx}%

\IEEEoverridecommandlockouts

\title{\LARGE \bf NDBench: Benchmarking Microservices at Scale}

\author{
    \IEEEauthorblockN{Ioannis Papapanagiotou and Vinay Chella}
    \IEEEauthorblockA{Cloud Database Engineering \\
    Netflix \\
    Los Gatos CA 95032
    \\\{ipapapanagiotou, vchella\}netflix.com}
}

\begin{document}

\maketitle
\thispagestyle{empty}
\pagestyle{empty}

%


\begin{abstract}

Software vendors often report performance numbers for the sweet spot or running on specialized hardware with specific workload parameters and without realistic failures. Accurate benchmarks at the persistence layer are crucial, as failures may cause unrecoverable errors such as data loss, inconsistency or corruption. To accurately evaluate data stores and other microservices at Netflix, we developed Netflix Data Benchmark (NDBench), a Cloud benchmark tool. It can be deployed in a loosely-coupled fashion with the ability to dynamically change the benchmark parameters at runtime so we can rapidly iterate on different tests and failure modes. NDBench offers pluggable patterns and loads, support for pluggable client APIs, and was designed to run continually. This design enabled us to test long-running maintenance jobs that may affect the performance, test numerous different systems under adverse conditions, and uncover long-term issues like memory leaks or heap pressure. 

\end{abstract}

%
%


\section{Introduction}
\label{sec:intro}
Netflix runs thousands of microservices and thousands of backend data store instances on Amazons Web Services (AWS) to serve and store data for more than 120 million users. Our operational data scale encompasses petabytes of data, trillions of operations per day. and tens of millions of real-time, globally replicated, mutations per second. Netflix also runs in active-active mode. This means that services are stateless, all data/state is being asynchronously replicated by the data tier, and resources must be accessed locally within a region. One of the fundamental challenges in implementing Active-Active is the replication of user's data. Therefore, we are using data store solutions that support multi-directional and multi-datacenter asynchronous replication.


Due to this architecture, we cannot always characterize the load that microservices apply to our backend services. In particular, we cannot predict the load when a failure occurs in the data system or in one of the microservices or in the underlying cloud provider. For example, when an AWS region is experiencing an outage, large numbers of clients simultaneously failover to other regions. Understanding the performance implications of new microservices on our backend systems is also a difficult task. For this reason, we run regular exercises on a weekly basis that test Availability Zone outages, split brain situations (data replication getting queued up) or regional outage simulations. We also need a framework that can assist us in determining the behavior of our scalable and highly available persistent storage and our backend infrastructure under various workloads, maintenance operations, and capacity constraints. We also must be mindful of provisioning our clusters, scaling them either horizontally (by adding nodes) or vertically (by upgrading the instance types), and operating under different workloads and conditions, such as node failures, network partitions, etc.

As new distributed NoSQL data store systems appear in the market or pluggable database engines based on different algorithms like Log-Structured Merge-Tree (LSM) design \cite{rocksdb}, B-Tree, Bw- Tree \cite{levandoski2013bw}, HB+ Trie \cite{forestdb} etc. they tend to report performance numbers for the ``sweet spot``, and are usually based on optimized hardware and benchmark configurations. Being a cloud-native enterprise, we want to make sure that our systems can provide high availability under multiple failure scenarios, and that we are utilizing our resources optimally. There are many other factors that affect the performance of a database such as the hardware, virtualized environment, workload patterns, number of replicas etc. There were also some additional requirements; for example, as we upgrade our data store systems we wanted to integration test the distributed systems prior to deploying them in production. For other systems that we develop in-house (EVCache \cite{evcache} for key/value caching, Dynomite \cite{dynomite} for structured data caching) and for our contributions to Apache Cassandra, we wanted to automate the functional test pipelines, understand the performance of these systems under various conditions, and identify issues prior to releasing for production use or committing code to upstream. Hence, we wanted a workload generator that could be integrated into our pipelines prior to promoting an Amazon Machine Image (AMI) to a production-ready AMI. 

Finally, we wanted a tool that would resemble our production deployment. In the world of microservices, a lot of business process automation is driven by orchestrating across services. Traditionally, some of the benchmarks were performed in an ad-hoc manner using a combination of APIs, making direct REST calls in each microservice, or using tools that directly send traffic to the data stores. However, most benchmark tools are not integrated as microservices and do not have out of the box cluster management capabilities and deployment strategies. 

We looked into various benchmark tools as well as REST-based performance tools, protocol specific tools or tools developed for transactional databases to extend. We describe our related work research in section \ref{sec:bwwork}. While some tools covered a subset of our requirements, we were interested in a tool that could achieve the following:
\begin{itemize}
\item Single benchmark framework for all data stores and services
\item Dynamically change the benchmark configuration and workload while the test is running.
\item Test a platform along with production microservices
\item Integrate with fundamental platform cloud services like metrics, discovery, configurations etc.
\item Run for an unlimited duration in order to test performance under induced failure scenarios and long running maintenance jobs such as database anti-entropy repairs.
\item Deploy, manage and monitor multiple instances from a single entry point.
\end{itemize}

For these reasons, we created Netflix Data Benchmark (NDBench). NDBench allows us to run infinite horizon tests that expose failure conditions such as thread/memory leaks or sensitivity to OS disk cache hit ratios from long running processes that we develop or use in-house. At the same time, in our integration tests we introduce failure conditions, change the underlying variables of our systems, introduce CPU intensive operations (like repair/reconciliation), and determine the optimal performance based on the application requirements. Finally, we perform other activities, such as taking time based snapshots or incremental backups of our databases. We want to make sure, through integration tests, that the performance of these system is not affected. 

We incorporated NDBench into the Netflix Open Source Software (OSS) ecosystem by integrating it with components such as Archaius for configuration management \cite{archaius}, Spectator for metrics \cite{spectator}, and Eureka for service registration \cite{eureka}. However, we designed NDBench so that these libraries are injected, allowing the tool to be ported to other cloud environments, run locally, and at the same time satisfy our Netflix OSS ecosystem users.  We have been running NDBench for almost three years, having validated multiple database versions, tested numerous NoSQL systems running on the Cloud, identified bugs, and tested new functionalities.
\section{Architecture}
\label{sec:arch}

NDBench has been designed as a platform that teams can use by specifying workloads and parameters. It is also adaptable to multiple cloud infrastructures. NDBench further provides a unified benchmark tool that can be used both for databases as well as other backend systems. The framework consists of three components:
\begin{itemize}
  \item Plugins: The API where data store plugins are developed 
  \item Core: The workload generator
  \item Web: NDBench UI and web backend
\end{itemize}

\begin{figure}[!t]
\centering
\includegraphics[width=0.45\textwidth]{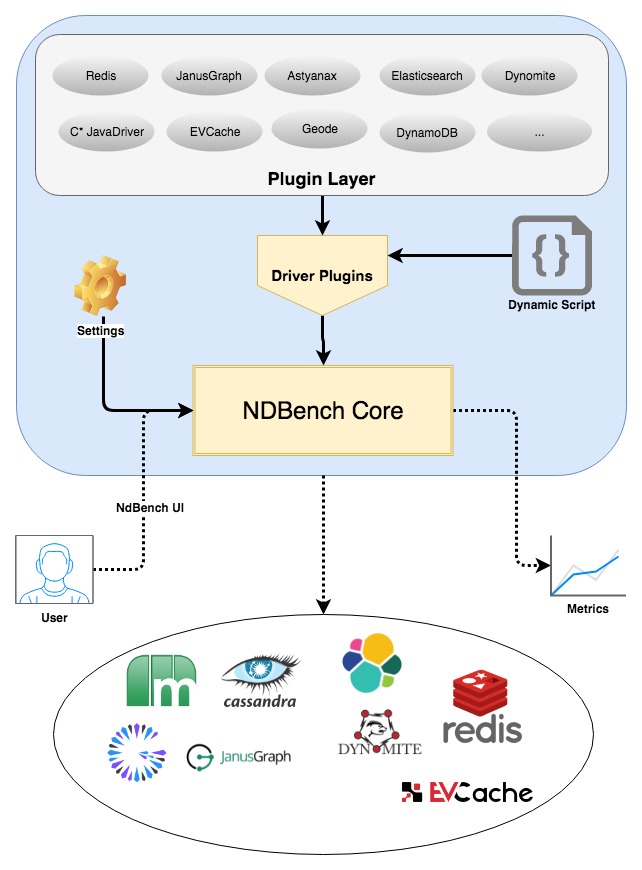}
\caption{NDBench Architecture}
\label{nd_arch}
\end{figure}

NDBench offers the ability to add client plugins for different caches and databases allowing users to implement new data store clients without rewriting the core framework. Adding a plugin mainly consists of two parts:

\begin{itemize}
  \item An implementation of an interface that initializes the connection to the system under investigation and performs reads/writes
  \item (Optional) Adding a configuration interface for the system under investigation.
\end{itemize}
All remaining benchmark components are handled by NDBench core functionality such as metrics, multi-threaded load generation etc. At the time that this paper was written, NDBench supports a number of drivers including the Datastax Java Driver using the Cassandra Query Language (CQL), Thrift support for Apache Cassandra \cite{astyanax}, Elasticsearch REST and native APIs for document storage, Redis, Memcache clients for in-memory storage (and their corresponding Netflix wrappers for EVCache \cite{evcache} and for Dynomite \cite{dynomite}), Amazon Web Services DynamoDB \cite{decandia2007dynamo}, Apache Janusgraph plugin with support of Tinkerpop for graph frameworks, and Apache Geode for data management. NDBench was open sourced a year ago \cite{ndbench}, hence some of the plugins have been developed by Netflix engineers and others from NDBench users/committers.

A unique ability of NDBench is that a plugin is not confined by the limits of a single data system. For example, a plugin can contain multiple drivers. For example, we may want to check how a microservice performs when adding a logic for a side-cache like Memcached, and a database like Cassandra. More complicated systems that exist nowadays may support Change Data Capture (CDC), like a primary data source like Cassandra and a derived data source like Elasticsearch for advanced indexing or text search. One can therefore extend plugins to contain multiple client drivers based on the corresponding business logic.

\begin{figure*}[!t]
\centering
\includegraphics[width=0.7\textwidth]{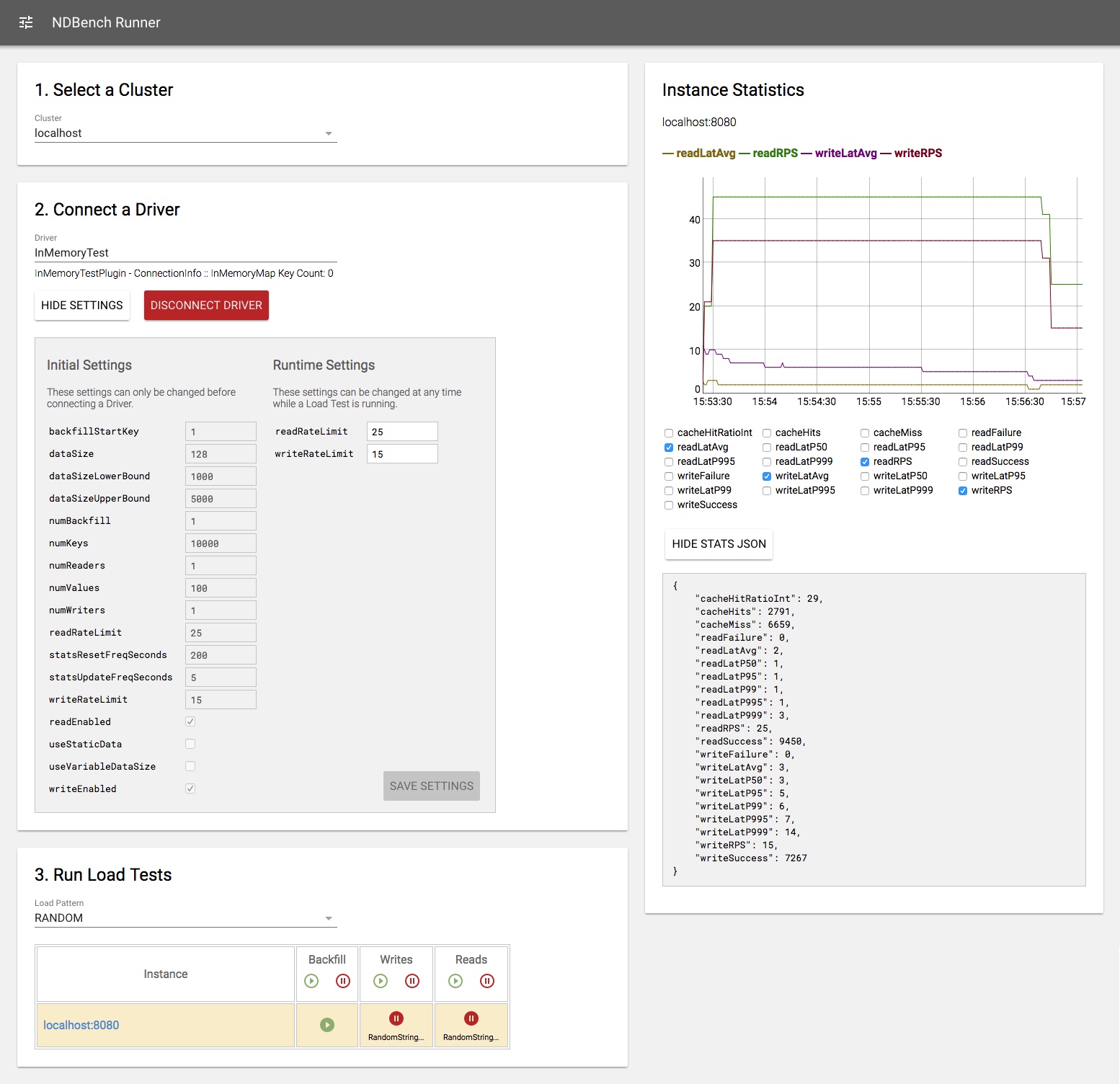}
\caption{NDBench User Interface}
\label{nd_ui2}
\end{figure*}

The NDBench components are shown in Fig. \ref{nd_arch}. NDBench-core is the core component of the software. The client drivers are embedded inside the plugins. NDBench loads the plugins at runtime and then the plugins are responsive to send traffic directly to the systems under investigation. NDBench-core also collects user parameters through a RESTful interface. These parameters can either be provided by a UI or by simple REST calls made via the command line. This allows for a microservice deployment of NDBench to receive input from other services. NDBench also receives other input parameters such as the data store cluster to communicate, port number etc. These parameters are then passed to the client driver. Hence, when we deploy on the cloud, we do not need to specify multiple hostnames and port numbers, but rather a single name on the service registration. NDBench will then use the idiosyncrasies of each client driver to determine what is the proper strategy (token aware, round robin, etc.) to send requests to the data store cluster. At Netflix, we isolate AWS Regions and Availability Zones by forcing clients to have region affinity. In case of a failure, clients fail over to alternative zones but they never cross the regional boundaries. This achieves the right trade-off of latency and availability for us. As such, the only traffic that gets replicated across regions is done by our data store systems. 

The NDBench UI is shown in Fig. \ref{nd_ui2}. The user starts different operations on a per node basis of the NDBench cluster either through the UI or programmatically through REST calls. Backfill is used to pre-load the data store so that the test can begin in a realistic state. This avoids the ramp up time needed when simulating workloads. The user can then select to either start writes, reads, or both concurrently. Unlike other benchmark tools, NDBench allows the reads to contain any type of read operations that the user has embedded in the read port of the client plugin. As a result, reads can contain both point queries as well as range queries. Similarly, writes can contain different write operations. This allows for more complex, production like, workloads.

\subsection{Parameters}
\subsubsection{Core Parameters}
NDBench provides a variety of input parameters that are loaded dynamically and can change during the workload test. This is a core functionality of NDBench. It allows the test to be modified at runtime which tests how the overall architecture is performing under changing conditions. More crucially, as our architecture consists of thousands of microservices, the complexity and variability of our loads has grown and required this dynamic capability. The organic growth of our subscriber base across the globe in combination with our batch workloads leads to peaks and troughs in different time frames. Our philosophy remains unchanged around injecting failures into production to ensure our systems are fault-tolerant \cite{basiri2016chaos}. Running, configuring and tweaking NDBench while our systems run has been a major requirement in developing NDBench. NDBench allows the following parameters to be configured on a per node basis:

\begin{itemize}[itemsep=0mm]
 \item \texttt{numKeys}: the sample space for the randomly generated keys
 \item \texttt{numValues}: the sample space for the generated values
 \item \texttt{dataSize}: the size of each value
 \item \texttt{numWriters/numReaders}: the number of threads per NDBench node for writes/reads
 \item \texttt{writeEnabled/readEnabled}: a boolean to enable or disable writes or reads
 \item \texttt{writeRateLimit/readRateLimit}: the number of writes per second and reads per seconds
 \item \texttt{userVariableDataSize}: a boolean to enable or disable the ability of the payload to be randomly generated.
\end{itemize}

\begin{figure*}[!t]
\centering
\includegraphics[width=0.6\textwidth]{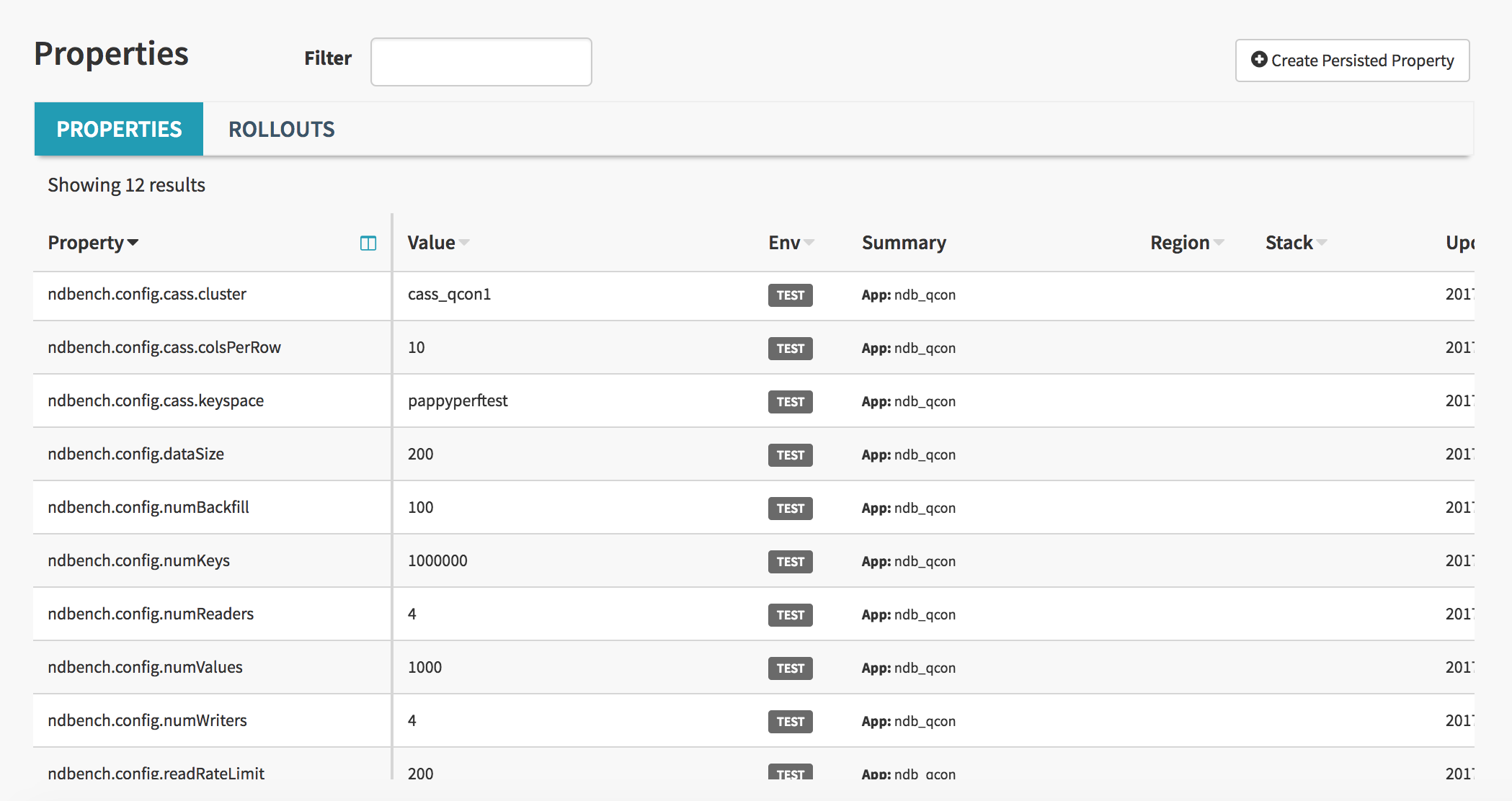}
\caption{Integrating Runtime Configurations with Continuous Delivery}
\label{nd_fp}
\end{figure*}

NDBench enables the user to plug in any properties framework. Hence, parameters can be provided through a local file, a remote database, or by any other property framework. At Netflix, we have integrated the properties of NDBench with our continuous delivery system. This is shown in Fig. \ref{nd_fp}. Hence configuring those properties can be done by a single entry point. Given the dynamic nature of some of these properties, we can observe the output directly on the NDBench UI.

\subsubsection{Plugin Parameters}
NDBench allows the user to programmatically change the parameters of the plugin. Hence, as long as the client driver provides the ability to change parameters one can do that through the same code base instead of having to have an external configuration, a sidecar or going through an additional UI to configure a table or data store system. For example, when we want to evaluate our Memcache infrastructure we are interested in defining the effect of changing the TTL of the data and the size of our caching layer in relation to the cache hit ratio. We were interested in doing that through the single code. In the case of DynamoDB, we were interested in defining the table, schema, write and read capacity units, as well as whether we are interested in consistent reads.

\subsection{Workload configuration}
NDBench offers plugable load tests. This provides the ability to our developers to create new tests that represent new or existing use cases. To assist with using our developers that do not have strict requirements on traffic, NDBench has built-in the following distributions:
\begin{itemize} 
 \item \textbf{uniform}: choose a key randomly for the duration of the test, hence equally distributing the load across all the database shards.
 \item \textbf{zipfian}: choose a key based on the Zipfian distribution. This allows to distinguish between hot and cold keys, as the hot keys would have higher chance of being selected.
 \end{itemize}

Additionally, NDBench is not limited to these distributions. It allows users to deploy new distributions (including Binomial, Geometric, Hypergeometric, Pascal, Poisson, and Uniform) that match their workloads better. However, our experience shows that a single distribution may not properly emulate a production environment, which may benefit from temporal and spatially local data. 
 
The locality is important in our systems as they exercise the embedded caching of the data store system (OS page cache, block cache, or heap), deployments of server side caching solutions like Redis and Memcached as well as the disk's IOPS (Input/Output Operations Per Second). Hence, we have added a new form of traffic generator called \textit{sliding window}. The \textit{sliding window} test leverages one of the aforementioned distributions to generate data inside a sliding window. This test provides valuable data about how datastores react to changing key distributions over time.

\subsection{Deployment}
In a public cloud deployment, service registry and discovery is crucial in order to communicate with a cluster. A service registry allows fast rollback of versions in case of problems avoiding the re-launch of 100's of instances which could take a long time. Additionally, in rolling pushes, it avoids the propagation of a new version to all instances in case of problems.
Similarly it allows us to take instances of our Cassandra \cite{lakshman2010cassandra} fleet out of traffic for maintenance. NDBench receives information from our AWS service registry named Eureka through a REST call \cite{eureka}. The request is formulated leveraging the cluster name the user has used in the initialization process. The response is provided in JSON format. NDBench parses the JSON that contains the instance ids, port number, the status of the node based on the internal healthcheck, and other related information about the data store instances. This process allows NDBench to be adaptable to any public cloud platform and at the same time embed the functionality of service discovery. When a node fails, NDBench uses the underlying logic of the driver to failover traffic. For example, in our Cassandra, Dynomite and EVCache deployments the client would failover to another node in the same region.

NDBench provides a multitude of ways to configure the properties of the cluster or the local deployment. The configuration of NDBench follows the principles of Apache's Commons Configuration Library based on the integration of the Netflix Java configuration management API called Archaius \cite{archaius}.
The Archaius integration allows a number of features such as dynamic and type specific properties, high throughput and thread safe configuration operations, a polling framework that allows NDBench to obtain property changes to a configuration source, a callback mechanism that gets invoked on effective/``winning`` property mutations (in the ordered hierarchy of Configurations) and a  Java Management Extensions (JMX) Managed Bean (MBean) that can be accessed via JConsole to inspect and invoke operations on properties. 

NDBench provides a number of deployment strategies. A local deployment, in which the properties are fed from a local configuration file, or various modes of Cloud deployment. NDBench as a cloud microservice can leverage deployment strategies such as Blue-Green (or Red-Black as we call it at Netflix),  ``highlander'' (``there can only be one!'') or canary deploys. It can also be pinned in different clusters. The configuration of the NDBench properties can be easily managed through Archaius. For example, properties can be scoped on an instance basis, rack, datacenter etc. At the same time, NDBench workload can be generated on a cluster-wide basis or in individual instances through the central UI.

\subsection{Auto-tuning}
At Netflix, each microservice generates different types of loads and the load is not always known before we create the data store cluster. Hence, we had to advise our engineers about the size of the cluster and the scalability limitations  - for example in Cassandra we double our clusters if we want to scale horizontally to achieve equal distribution of the traffic. At the same time, we do not want to overspend by creating artificially large clusters.

In the past, the process of figuring out how much load a cluster could handle was manual. We would run NDBench with different throughput and load patterns, data store cluster instance types, and nodes. We would find the golden spot at which the cluster was not too overloaded yet also not too underutilized. We also have to budget Chaos \cite{basiri2016chaos} exercises in which the traffic shifts from one region to another region. This manual pain motivated the auto-tuning feature.

The auto-tuning feature automates the aforementioned process by gradually increasing request load, until some percentage of SLA violations occur, such as percentile latencies exceeding a set limit, mutations dropped by a database or index document failures in a document store. NDBench performs an exponential backoff algorithm and steps down the load based on preset $\%$ of SLA violation. This is needed because in some scenarios the data stores may be only slightly overloaded and may have passed the threshold at which they can continue to operate within SLA.

\section{Use Cases}
\label{sec:eval}
NDBench is used to evaluate multiple data stores and microservices across multiple hardware and software configurations. The pluggable nature of NDBench provides detailed performance analysis which allows Netflix to compare between and within deployments of diverse APIs.



Before each deployment of our data services, NDBench tests run to ensure that the service meets the product team's latency and availability requirements under adverse conditions.

\subsection{Comparing Cloud Instances}
NDBench is a fundamental tool for benchmarking new hardware instance types. It allows us to look across different dimensions on how the hardware performs under different types of workloads. A recent example is in 2017 AWS announced the release of the i3 instance family. These instances are equipped with NVMe SSD storage, and can deliver several millions IOPS at a 4KB block size and tens of GB/second of sequential disk throughput. At Netflix, we run Apache Cassandra \cite{lakshman2010cassandra} on AWS EC2 instances. Cassandra is the main data store supporting thousands of microservices and more than 125 million subscribers at Netflix. It serves as a solid foundation for Netflix's globally replicated dataset, bringing non-linear internet TV to customers around the world. With hundreds of clusters, tens of thousands of nodes and petabytes of data, Cassandra serves several millions of operations/sec with multiple nines of availability at the storage layer.

\begin{figure*}[!t]
\centering
\includegraphics[width=0.85\textwidth,height=0.25\textheight]{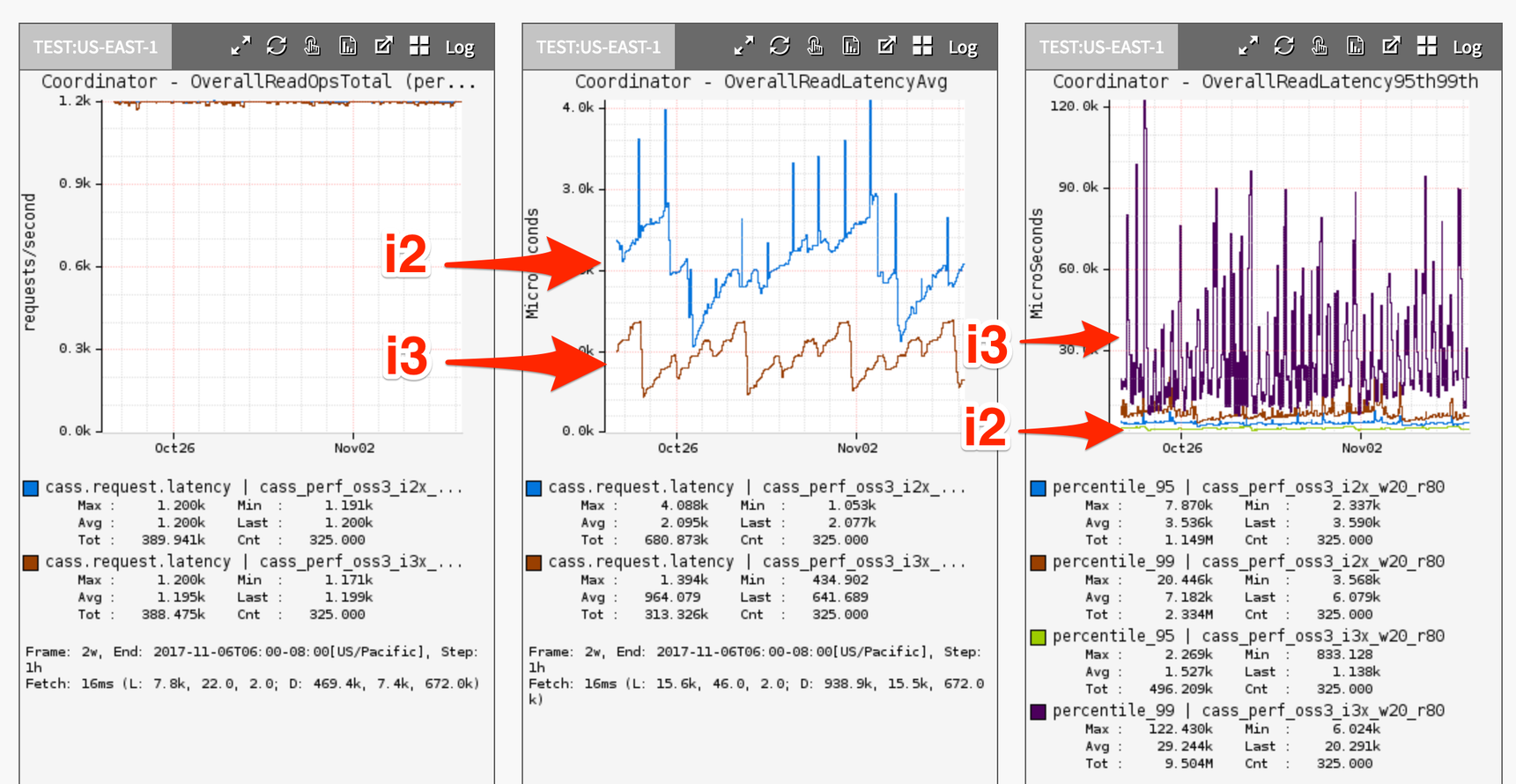}
\caption{Cassandra Performance on EC2 i2 vs i3 instances}
\label{i2_vs_i3}
\end{figure*}

Given the nature of our Cassandra and RocksDB \cite{rocksdb} deployment, i.e. LSM trees, these instances seemed to be a perfect match for Netflix data storage. We therefore used NDBench to certify them as production ready. In the first iteration, we ran a 80/20 R/W NDBench generated workload with point queries. We used two NDBench clusters, each one sending traffic to a different Cassandra cluster. One benchmark was run against i2 instances that we have been using in our production systems, and the other ran on newly released i3 instances. In the first iteration, we observed that the throughput in both case were the same and the average latencies were lower on i3 instance family. However, the 99th percentile read latencies were significantly higher and unpredictable on the i3 instances (as high as 100ms), which is shown in Fig. \ref{i2_vs_i3}. This variability far exceeds our production latency requirements. We determined these slow reads happened because the version of the Linux kernel we ran had no disk scheduler for multiqueue drives, so when large background OS page cache flushes saturated the i3's drives with many write operations, latency sensitive reads of mmaped files would queue behind them. 

\begin{figure*}[!t]
\centering
\includegraphics[width=0.85\textwidth,height=0.20\textheight]{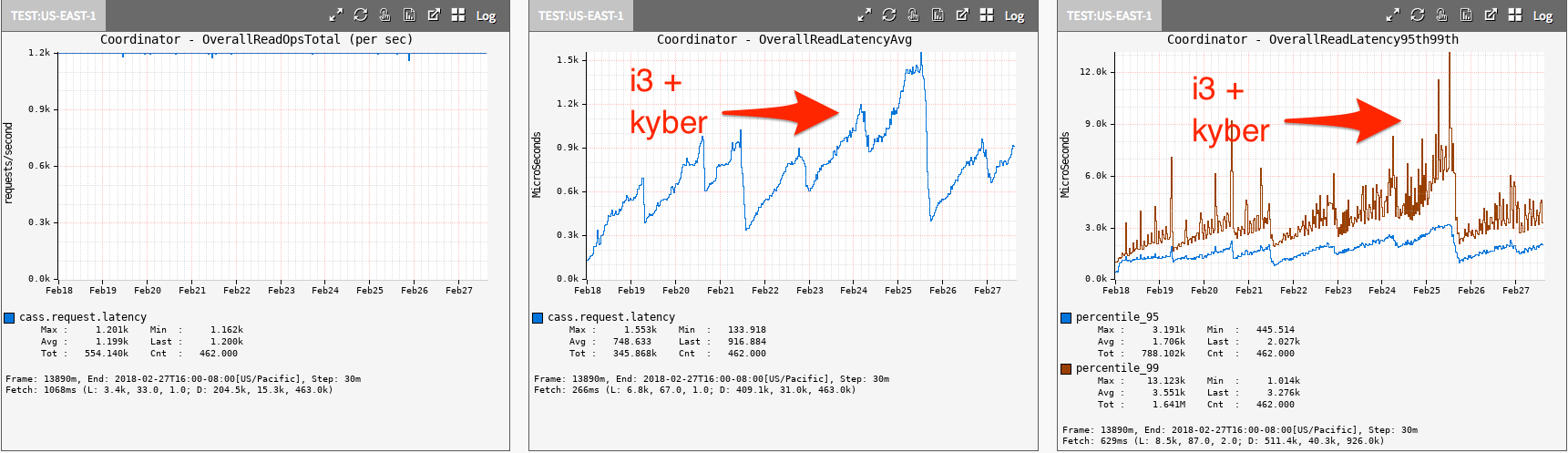}
\caption{Cassandra Performance on EC2 i3 instance with Kyber IO scheduler}
\label{i3_kyber}
\end{figure*}

After performing a detailed performance analysis we upgraded to Linux 4.13, which includes the Kyber \cite{kyber} multiqueue I/O scheduler. This scheduler is able to fairly prioritize reads with writes, and after switching to it, the 99th percentile read latency dropped to acceptable levels as reported by NDBench as seen in Fig. \ref{i3_kyber}. 

Without NDBench's ability to run complex mixed mode workloads over long periods of time, we would not have seen these unacceptable read latencies \textit{caused by periodic background writes}.

\subsection{Configurations}
We furthermore use NDBench to evaluate different configurations of data stores. For example, many storage and streaming systems are written in Java and since Java 1.8, there are multiple garbage collectors one can choose from. The Concurrent-Mark-Sweep (CMS) Collector employs multiple threads (``concurrent'') to scan through the heap (``mark'') for objects that are unused and therefore can be recycled (``sweep''). The Garbage-First collector (G1) introduced in JDK 7 update 4 was designed to better support heaps larger than 4GB. The G1 collector utilizes multiple background threads to scan through the heap that it divides into regions, spanning from 1MB to 32MB (depending on the size of your heap). The G1 collector is geared towards scanning those regions that contain the most garbage objects first, giving it its name (Garbage-First). 

\begin{figure*}[!t]
\centering
\includegraphics[width=0.85\textwidth,height=0.3\textheight]{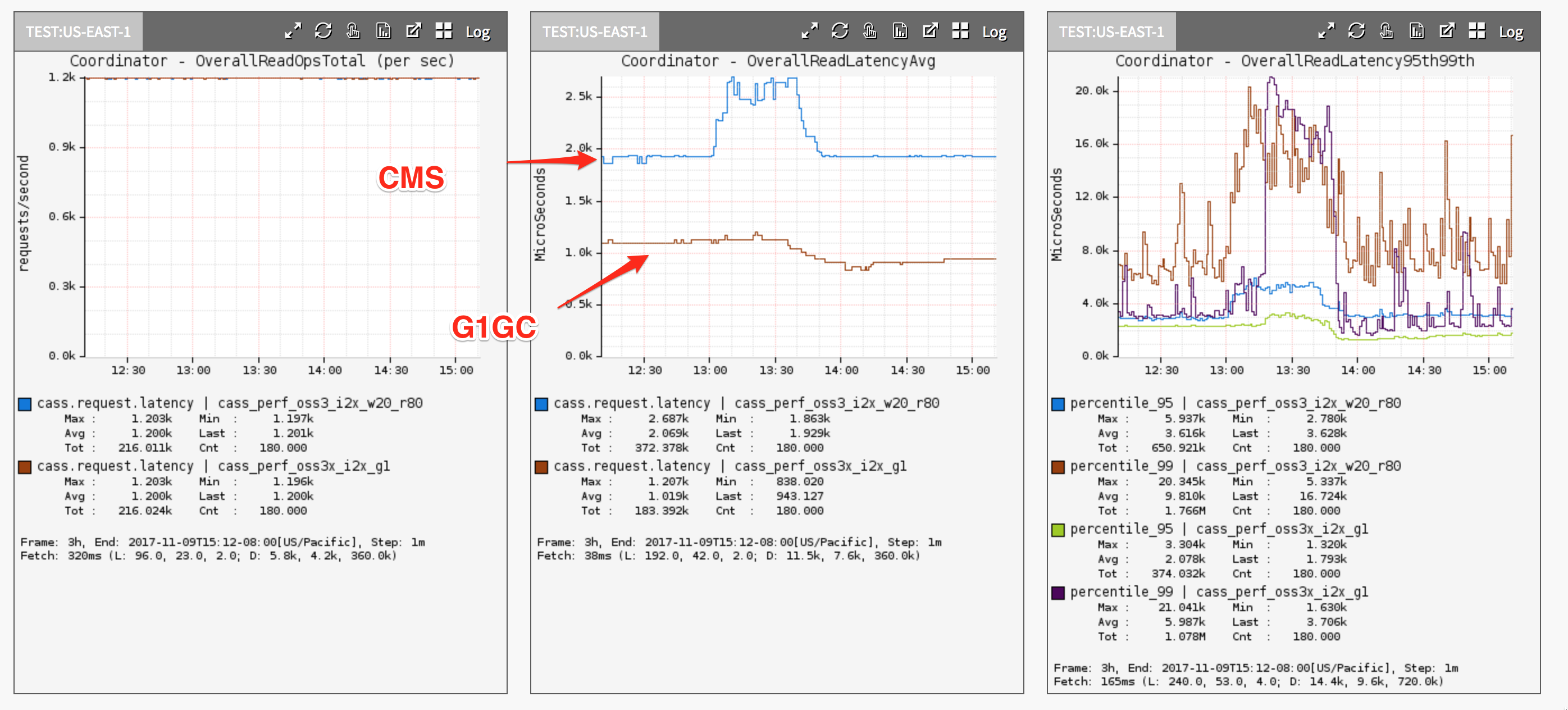}
\caption{Cassandra Performance on CMS vs G1GC Garbage Collectors}
\label{CMSvsG1GC}
\end{figure*}

We want to test the performance of G1GC prior to changing production configurations. In Fig. \ref{CMSvsG1GC}, we observe the comparison under an NDBench generated 80/20 R/W point query workload with randomly generated payloads of 1KB. We see that for the same throughput G1GC has almost half the average latencies (1ms vs 2ms) at the coordinator level as well as half the 99th percentile read latencies (4ms vs 8ms). This data helps inform future production deployments of Cassandra.

\subsection{AMI Certification}
NDBench's API allows for further automation of benchmarking. At Netflix we include it as part of our CD pipelines that deploy new AMIs to our fleets. The process consists of a number of tests including integration and performance testing. This way every AMI that we deploy is rigorously certified using NDBench generated load.

To achieve this, we integrated NDBench with Spinnaker, our multi-cloud continuous delivery platform \cite{spinnaker}. In Spinnaker, pipelines are the key deployment management construct. They consist of a sequence of actions, known as stages. One can pass parameters from stage to stage along the pipeline. For example, start a pipeline manually, or start by automatic triggers, such as a Jenkins job, a  time-based job scheduler (e.g. cron), or a stage in another pipeline. Pipelines can be configured to emit notifications at various points during the pipeline execution (such as pipeline start/complete/fail), by email, SMS or any chat platform.

\begin{figure}[!t]
\centering
\includegraphics[width=0.5\textwidth]{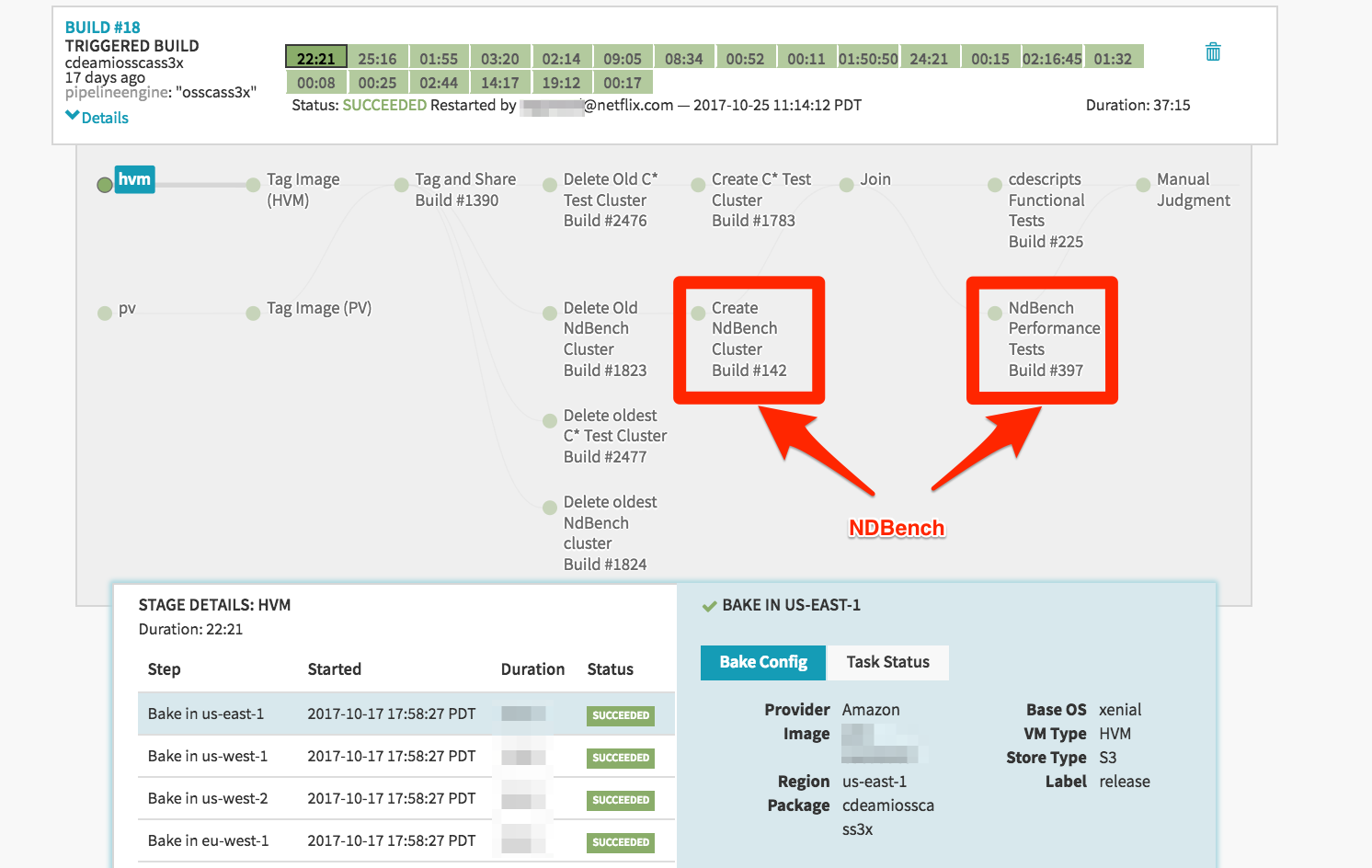}
\caption{AMI Certification Pipeline}
\label{eval_spin}
\end{figure}

We designed pipelines in Spinnaker and integrated NDBench into them, as shown in Fig. \ref{eval_spin}. Fig. \ref{eval_spin} shows the bake-to-release lifecycle of an AMI loaded with our database and the corresponding supporting frameworks such as sidecars, automation scripts etc. We initially bake an AMI with our data store, such as Cassandra. We then clean the environment so there are no outstanding databases or NDBench clusters with a similar name. The automation then creates a new NDBench cluster with the corresponding plugin and a database cluster with the newly baked AMI.

We then leverage a script that starts the load test initiated through a Jenkins job. While the test is running, we automatically perform different types of functional tests. The functional tests include restarting and terminating one or multiple nodes while the test is running, as well as testing some other database related jobs like anti-entropy repair. 

We finally review the results and make the decision on whether to promote an “Experimental” AMI to a “Candidate”. We use similar pipelines for most of our data store and caching systems such as Cassandra, Dynomite, Elastic Search, EVCache, testing out the replication functionalities with different client-side APIs. Passing the NDBench performance and functional tests means that the AMI is ready to be used in the production environment. 

\subsection{Database Auto-scaling}
\begin{figure}[!t]
\centering
\includegraphics[width=0.5\textwidth]{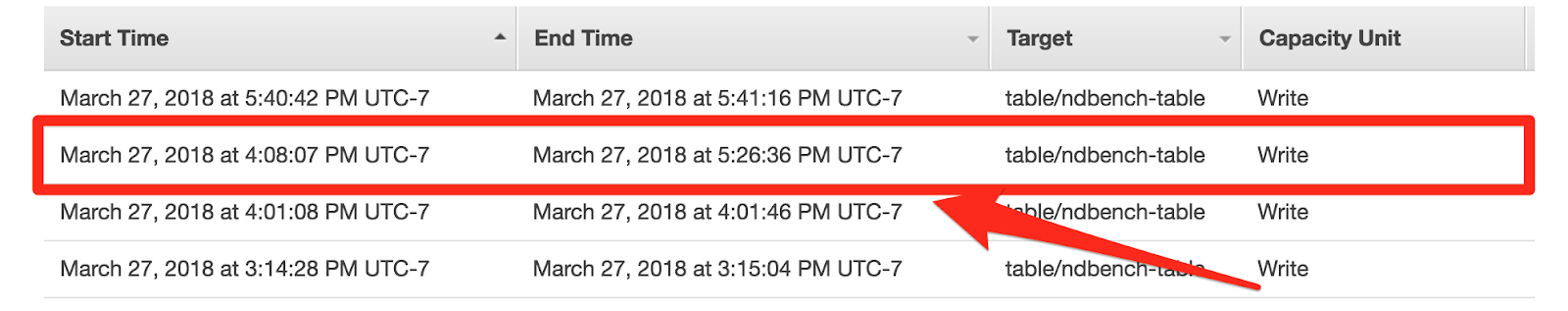}
\caption{DynamoDB Autoscale}
\label{ddb_auto}
\end{figure}

One of the plugins that NDBench support is AWS's DynamoDB.
DynamoDB works based on a provisioned capacity model and has an integrated auto-scaling feature. Auto-scaling helps with automating the capacity management for the tables and global secondary indexes. In DynamoDB, the user can set the desired target utilization and the system will adjust the provisioned capacity accordingly. As part of our evaluation of different use cases we wanted to be mindful of how we design some of microservices that use DynamoDB as their data store. More specifically, we wanted to test how fast a table can be scaled up (or down) and whether we are going to see any server-side issues if we ramp up the traffic while the service is running. In our experiments we used 100,000 Writes Per Second and increased the write load to 200,000 Writes Per Second without pre-splitting the partitions, hence forcing the worst case behavior. NDBench allowed us to see how the auto-scaling worked in DynamoDB. We observed that auto-scaling was very efficient and in most cases would complete in a few minutes to seconds and with a worst case scenario about 1h as shown with the red arrow in Fig. \ref{ddb_auto}.

\subsection{Write-through Cache}
\begin{figure}[!t]
\centering
\includegraphics[width=0.5\textwidth]{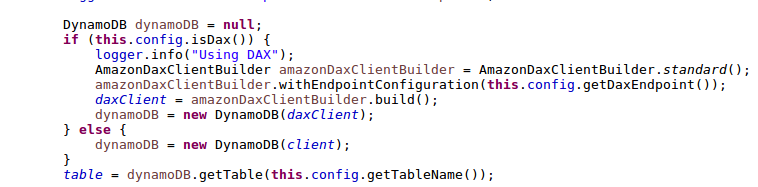}
\caption{Code snippet enabling DAX in DynamoDB Client API}
\label{ddb_code}
\end{figure}

Amazon DynamoDB Accelarator (DAX) is a write-through and read-through cache. Different from a side-cache, in which the application needs to have a different application logic to populate items to the cache and different logic to read and write to the data store, a write through cache allows to write and read data by a single API. A write/read through cache intercepts the requests and writes directly to the cache and then populates the database. As we wanted to evaluate this feature, we did not have to add a new DynamoDB plugin at NDBench. Instead, we created a simple dynamic property that would switch dynamically in the background between leveraging the DAX endpoint or hitting directly the DynamoDB service. NDBench allowed us to do that with a single $If$ statement as shown in \ref{ddb_code}. In other words, $config.dax()$ was a Boolean variable that would allows us to choose between which endpoint to use. 

\begin{figure}[!t]
\centering
\includegraphics[width=0.5\textwidth]{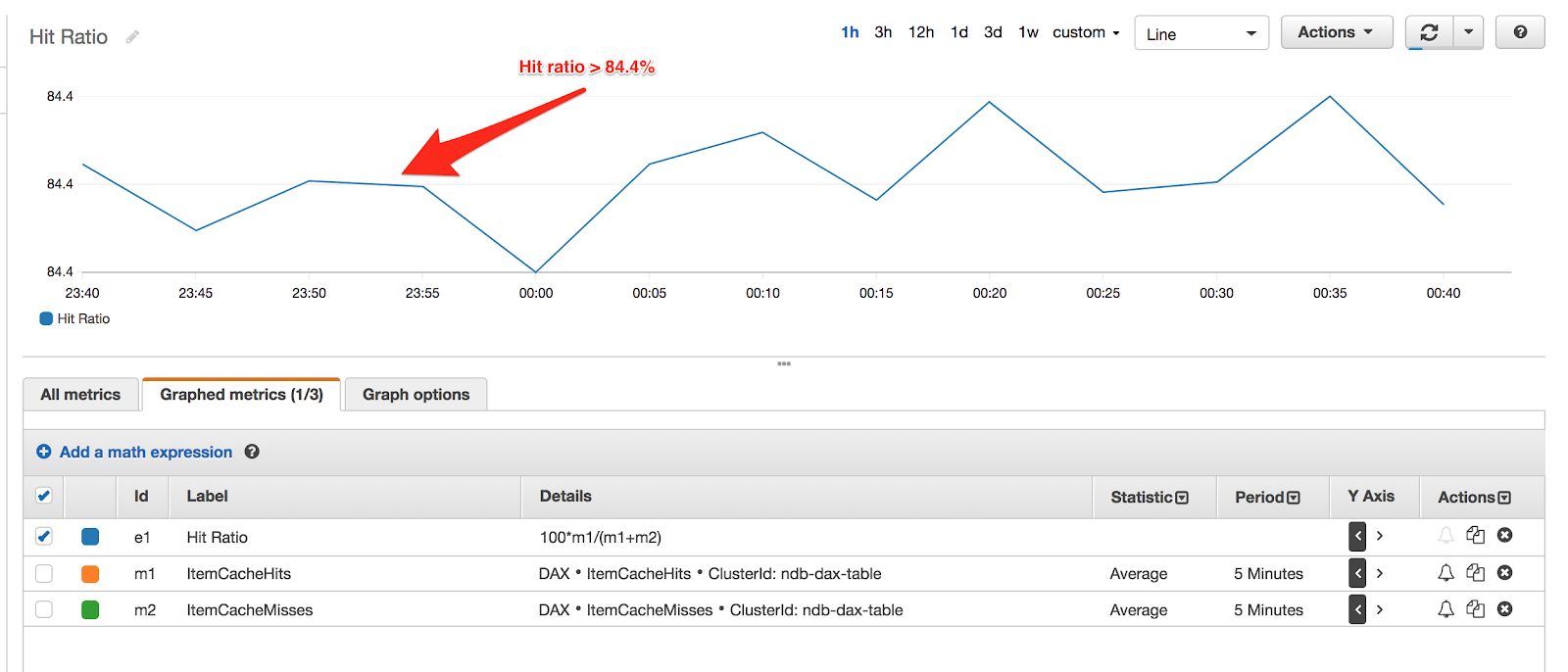}
\caption{DynamoDB Latency without DAX}
\label{ddb_dax_hitratio}
\end{figure}

NDBench support for the `sliding window` test allowed to test temporal and space locality in the way we generate data. This means that we could dynamically control the hit ratio of the cache, as shown in \ref{ddb_dax_hitratio}. By controlling the hit ratio above $>80\%$ would allow us to control the latencies of the write through cache in a single a run.

\begin{figure}[!t]
\centering
\includegraphics[width=0.5\textwidth]{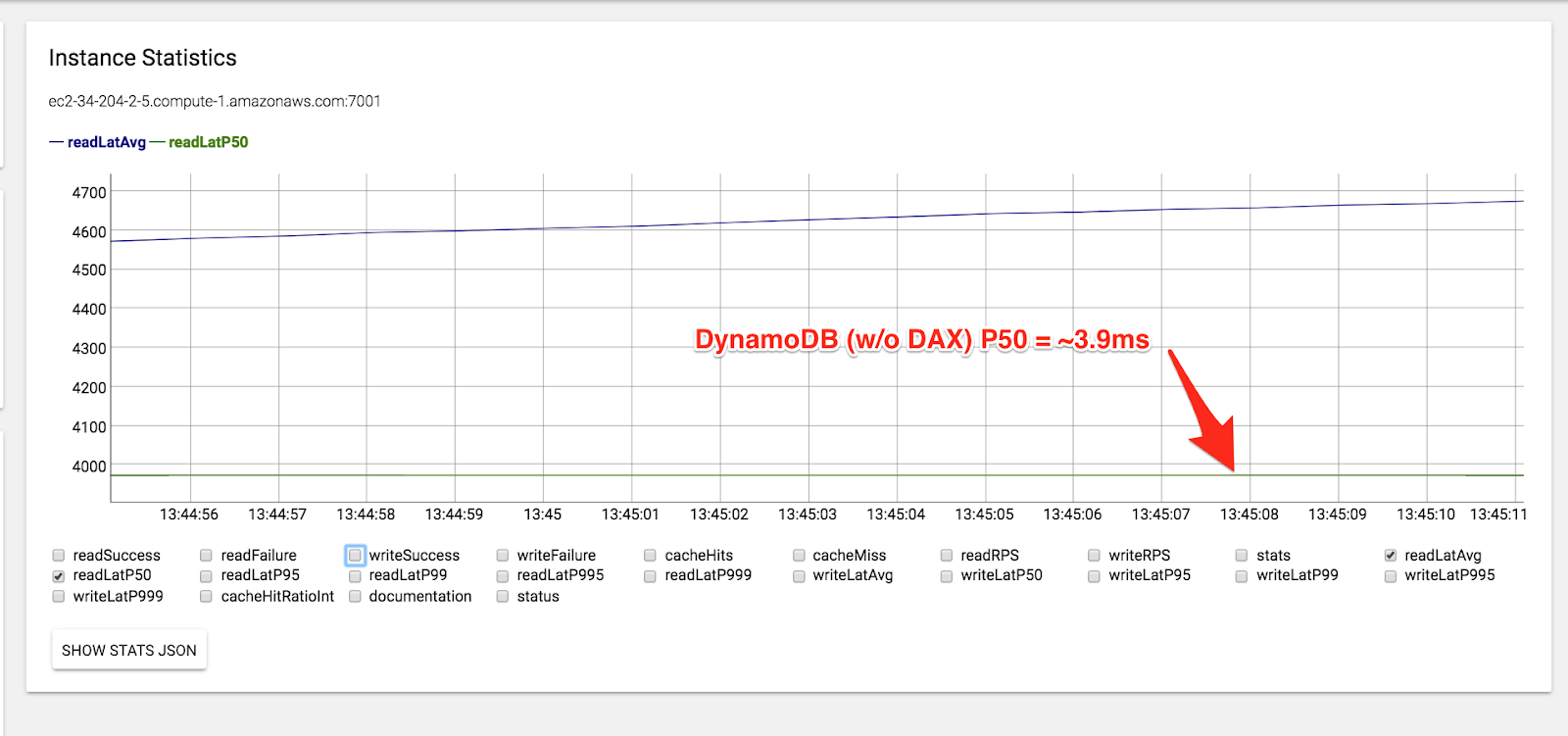}
\caption{DynamoDB Latency without DAX}
\label{ddb_no_dax}
\end{figure}

\begin{figure}[!t]
\centering
\includegraphics[width=0.5\textwidth]{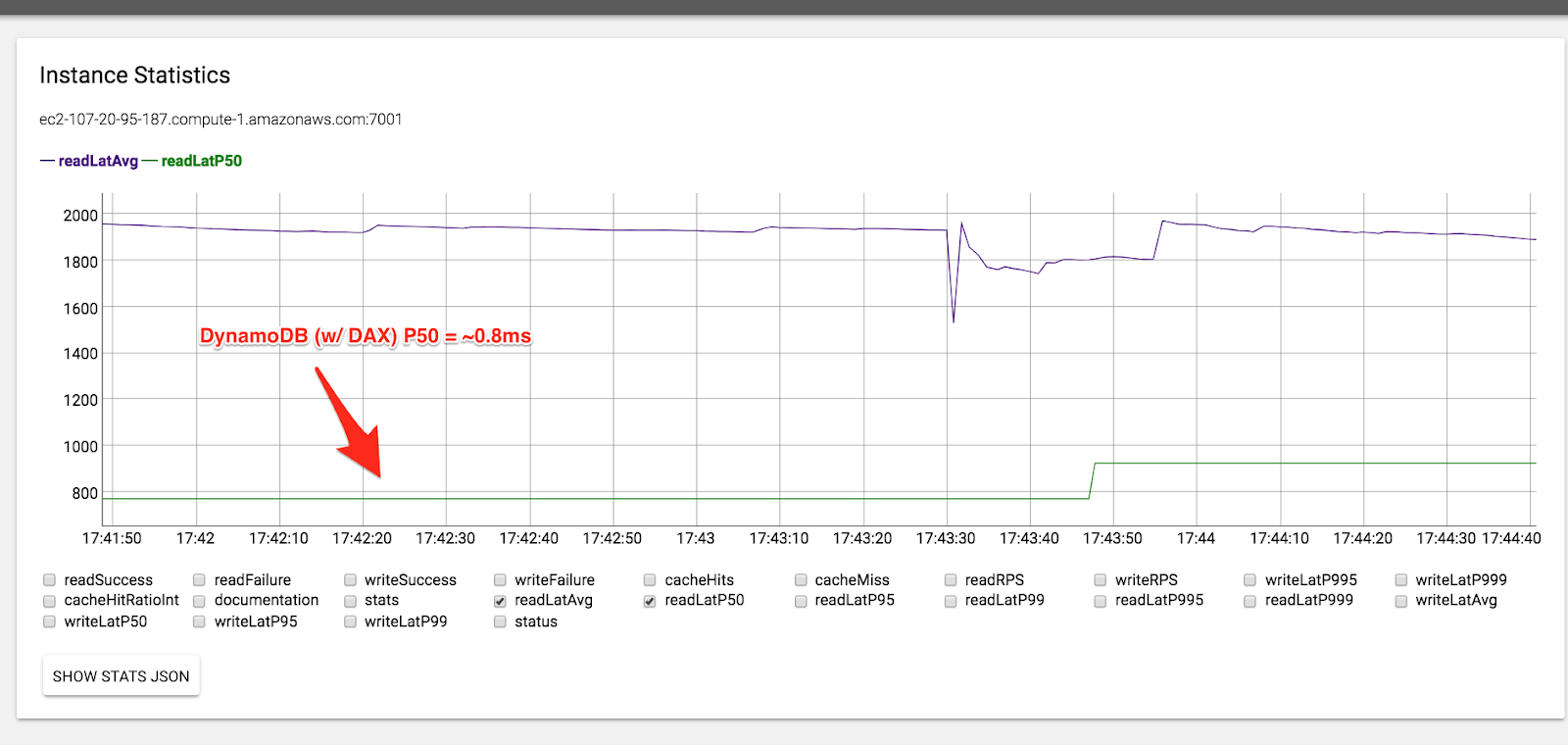}
\caption{DynamoDB Latency with DAX}
\label{ddb_dax}
\end{figure}

In Fig. \ref{ddb_no_dax}, we can see from the central NDBench UI the latencies as reported by the client without having to develop an external tool to monitor the latencies as NDBench has inherit support of showcasing the application latencies. At the same time, we could see different statistics in parallel by just clicking which statistics we want to showcase in the UI. In Fig. \ref{ddb_dax}, we were able to see a latency a drop in P50 latency from 3.98ms to 0.79ms. 

\section{Related Work}
\label{sec:bwwork}

\begin{table*}[!htbp]
\centering

\begin{tabular}{| l | c | c |}
\hline
  Features & YCSB & NDBench \\
\hline
  Multi-driver support & Yes & Yes \\
  Extensible Workload Generator & Yes & Yes \\
  Dynamic Parameters & N/A & Yes \\
  Central UI for Cluster Control & N/A & Yes \\
  Netflix OSS Cloud integration & N/A & Yes \\
  Benchmark Type & Finite Horizon & Infinite Horizon \\
  Horizontal Scale & Requires manual work & Yes \\
\hline
\end{tabular}
\caption{Comparison of benchmark frameworks}
\label{comp_table}

\end{table*}

One of the most well known and used frameworks for measuring the performance of several popular table stores is the Yahoo! Cloud Service Benchmark (YCSB) \cite{ycsb}. YCSB is fairly similar to our design in the sense that it provides an abstraction layer for adapting client APIs to specific data stores and it is probably the closest tool to compare with NDBench. It also provides the ability to gather metrics and generate a mix of workloads. We also looked into YCSB++ \cite{ycsb++}, an extension of YCSB that extends YCSB to multiple instances, provides eventual consistency measurements, and improves the performance and debugging of these features. However both YCSB and YCSB++ did not fit all of our requirements. YCSB does not allow us to configure cluster-wide parameters from a central location. Moreover, these tools perform finite horizon tests, and do not allow us to dynamically change the workloads or the data model while the benchmark was running. We summarized some of the differences in table \ref{comp_table}. A number of other related benchmark tools like Bibench \cite{bigbench} focused on big data use cases where in our use case we were not interested in a big data benchmark framework.

We also looked into frameworks specifically created for our data stores, like JMeter \cite{jmeter}, Cassandra-stress \cite{cassandra-stress} and Redis benchmark \cite{redis-benchmark}. JMeter \cite{jmeter} could provide us performance metrics for the Java process itself, whereas Cassandra-stress \cite{cassandra-stress} and Redis benchmark were limited to the corresponding data store systems C* and Redis respectively, and several other data store systems that we run, or plan to run in the future, at Netflix. Other REST based performance tools, like Gatling \cite{gatling}, would require that we run a REST service hence having an extra layer between the client our systems. Generally speaking each of these tools may work well for one data store, but we needed a consistent framework and platform for all data stores.

\section{Conclusions}
\label{sec:concl}
In this paper, we present NDBench, our cloud benchmarking framework. NDBench has the ability to dynamically change the benchmarking configurations while the test is running. This allows us to test leaks at the microservice layer, as well as performance under long-running background processes like data compactions, consistency repairs and node replacements. We have integrated NDBench with pluggable metrics, discovery, and configuration management platforms to make multi-cloud deployment possible. NDBench further has an integrated UI such that we can configure multiple benchmarks from a single interface, avoiding complicated configuration scripts.

We have used NDBench in a number of ways as a functional, integration and performance testing framework. We use NDBench in order to:
\begin{enumerate}
    \item Continuously test new data store code deployments under long running failure scenarios; 
    \item Test new backend systems and configurations;
    \item Evaluate hardware performance characteristics under complex and varied loads.
\end{enumerate}

NDBench was developed by the Cloud Database Engineering team at Netflix and therefore historically it has been used to evaluate data store systems. However, the pluggable architecture enables any microservice to microservice data interaction to be tested. There are a number of improvements we would like to make to NDBench, both for increased usability and supporting additional features. Some of the features that we would like to work on include performance profile management, automated canary analysis and dynamic load generation based on destination schemas.

\balance

\section{Acknowledgements}
We would like to acknowledge the contributions of Christopher Bedford for the auto-tuning feature and Kunal Kundaje for the UI design. Moreover, we are grateful to the open source committers for adding new client plugins as well as to our colleagues in the Cloud Database Engineering team for multiple contributions and discussion on NDBench. We appreciate the detailed contribution from Alexander Patrikalakis. We also appreciate the feedback on the paper from Lorin Hochstein and Joseph Lynch.
\bibliographystyle{abbrv}
\bibliography{citations}

\begin{thebibliography}{10}

\bibitem{gatling}
{Async Scala-Akka-Netty based Load Test Tool}.
\newblock \url{https://github.com/gatling/gatling}, 2017.
\newblock [Online; accessed 01-March-2018].

\bibitem{kyber}
{Kyber multiqueue I/O scheduler}.
\newblock \url{https://lwn.net/Articles/720071/}, 2017.
\newblock [Online; accessed 01-March-2018].

\bibitem{spinnaker}
{Netflix Spinnaker}.
\newblock \url{https://github.com/spinnaker/spinnaker}, 2017.
\newblock [Online; accessed 01-March-2018].

\bibitem{forestdb}
J.-S. Ahn, C.~Seo, R.~Mayuram, R.~Yaseen, J.-S. Kim, and S.~Maeng.
\newblock Forestdb: A fast key-value storage system for variable-length string
  keys.
\newblock {\em IEEE Transactions on Computers}, 65(3):902--915, 2016.

\bibitem{basiri2016chaos}
A.~Basiri, N.~Behnam, R.~de~Rooij, L.~Hochstein, L.~Kosewski, J.~Reynolds, and
  C.~Rosenthal.
\newblock Chaos engineering.
\newblock {\em IEEE Software}, 33(3):35--41, 2016.

\bibitem{ycsb}
B.~F. Cooper, A.~Silberstein, E.~Tam, R.~Ramakrishnan, and R.~Sears.
\newblock Benchmarking cloud serving systems with ycsb.
\newblock In {\em Proceedings of the 1st ACM Symposium on Cloud Computing},
  SoCC '10, pages 143--154, New York, NY, USA, 2010. ACM.

\bibitem{cassandra-stress}
Datastax.
\newblock Cassandra stress tool.
\newblock
  \url{https://docs.datastax.com/en/cassandra/2.1/cassandra/tools/toolsCStress_t.html},
  2016.

\bibitem{decandia2007dynamo}
G.~DeCandia, D.~Hastorun, M.~Jampani, G.~Kakulapati, A.~Lakshman, A.~Pilchin,
  S.~Sivasubramanian, P.~Vosshall, and W.~Vogels.
\newblock Dynamo: amazon's highly available key-value store.
\newblock {\em ACM SIGOPS operating systems review}, 41(6):205--220, 2007.

\bibitem{rocksdb}
S.~Dong, M.~Callaghan, L.~Galanis, D.~Borthakur, T.~Savor, and M.~Stumm.
\newblock Optimizing space amplification in rocksdb.
\newblock In {\em The biennial Conference on Innovative Data Systems Research},
  CIDR '17, 2017.

\bibitem{bigbench}
A.~Ghazal, T.~Rabl, M.~Hu, F.~Raab, M.~Poess, A.~Crolotte, and H.-A. Jacobsen.
\newblock Bigbench: Towards an industry standard benchmark for big data
  analytics.
\newblock In {\em Proceedings of the 2013 ACM SIGMOD International Conference
  on Management of Data}, SIGMOD '13, pages 1197--1208, New York, NY, USA,
  2013. ACM.

\bibitem{jmeter}
E.~H. Halili.
\newblock {\em Apache JMeter: A practical beginner's guide to automated testing
  and performance measurement for your websites}.
\newblock Packt Publishing Ltd, 2008.

\bibitem{lakshman2010cassandra}
A.~Lakshman and P.~Malik.
\newblock Cassandra: a decentralized structured storage system.
\newblock {\em ACM SIGOPS Operating Systems Review}, 44(2):35--40, 2010.

\bibitem{levandoski2013bw}
J.~J. Levandoski, D.~B. Lomet, and S.~Sengupta.
\newblock The bw-tree: A b-tree for new hardware platforms.
\newblock In {\em Data Engineering (ICDE), 2013 IEEE 29th International
  Conference on}, pages 302--313. IEEE, 2013.

\bibitem{archaius}
Netflix.
\newblock Archaius: Library for configuration management api.
\newblock \url{https://github.com/Netflix/archaius}, 2012.
\newblock [Online; accessed 01-March-2018].

\bibitem{astyanax}
Netflix.
\newblock {Astyanax}: Thrift client library for apache cassandra.
\newblock \url{https://github.com/Netflix/astyanax}, 2012.
\newblock [Online; accessed 01-March-2018].

\bibitem{evcache}
Netflix.
\newblock {EVCache}: A distributed in-memory data store for the cloud.
\newblock \url{https://github.com/Netflix/evcache}, 2013.
\newblock [Online; accessed 01-March-2018].

\bibitem{dynomite}
Netflix.
\newblock Dynomite: A generic dynamo implementation for different k-v storage
  engines.
\newblock \url{https://github.com/Netflix/dynomite}, 2014.
\newblock [Online; accessed 01-March-2018].

\bibitem{eureka}
Netflix.
\newblock Eureka: {AWS} service registry for resilient mid-tier load balancing
  and failover.
\newblock \url{https://github.com/Netflix/eureka}, 2014.
\newblock [Online; accessed 01-March-2018].

\bibitem{spectator}
Netflix.
\newblock Spectator: Client library for collecting metrics.
\newblock \url{https://github.com/Netflix/spectator}, 2016.
\newblock [Online; accessed 01-March-2018].

\bibitem{ndbench}
Netflix.
\newblock {Netflix Data Store Benchmark}.
\newblock \url{https://github.com/Netflix/ndbench}, 2017.
\newblock [Online; accessed 01-March-2018].

\bibitem{ycsb++}
S.~Patil, M.~Polte, K.~Ren, W.~Tantisiriroj, L.~Xiao, J.~L{\'o}pez, G.~Gibson,
  A.~Fuchs, and B.~Rinaldi.
\newblock Ycsb++: benchmarking and performance debugging advanced features in
  scalable table stores.
\newblock In {\em Proceedings of the 2nd ACM Symposium on Cloud Computing},
  page~9. ACM, 2011.

\bibitem{redis-benchmark}
S.~Sanfilippo.
\newblock {How fast is Redis?}
\newblock \url{https://redis.io/topics/benchmarks}, 2017.
\newblock [Online; accessed 01-March-2018].

\end{thebibliography}


\end{document}